\documentstyle[12pt]{article}
\newcommand{\dbs}{\renewcommand{\baselinestretch}{2.0}
\large\normalsize}

\def\csc{C$_{28}$C}

\def\cs{C$_{28}$}
\def\cd{(C$_{28})_2$}
\dbs
\begin{document}
\title{
Theoretical Study of Cubic Structures Based on
Fullerene Carbon Clusters:  C$_{28}$C and (C$_{28})_{2}$
}
\author{Linda M. Zeger, Yu-Min Juan and Efthimios Kaxiras\\
{\it Department of Physics and Division of Applied Sciences}\\
{\it Harvard University, Cambridge, MA 02138}\\
A. Antonelli\\
{\it Instituto de Fisica, Universidade Estadual de Campinas-Unicamp }\\
{\it Caixa Postal 6165, Campinas, Sao Paulo 13.081 Brazil}\\}


\maketitle
\begin{abstract}
We study a new hypothetical form of solid carbon \csc,
with a unit cell which is composed of the  \cs \ fullerene cluster
and an additional single carbon atom arranged in
the zincblende  structure.  Using {\it ab initio} calculations,
we show that this new form of solid carbon has lower energy than hyperdiamond,
the recently proposed form composed of \cs \ units in
the diamond structure.  To understand the bonding character of
of these cluster-based solids, we analyze
the electronic structure  of \csc \ and of hyperdiamond
and compare them to the electronic states of crystalline cubic diamond.
\end{abstract}

\dbs


\section{Introduction}

The \cs \ cluster has aroused considerable interest recently.
This unit is the smallest fullerene that has been produced in significant
quantities in experiments of laser vaporization of graphite \cite{guo}.
The structure of \cs \ \cite{kroto} is shown in Figure~\ref{strucc28}.
The twelve pentagons and four hexagons that comprise this structure
are arranged in a pattern that gives rise to three nearly tetrahedral
bond angles around four apex atoms (two such atoms are marked by A in
Figure~\ref{strucc28}).
Dangling $sp^{3}$ orbitals
on
the  four apex atoms
render this cluster chemically reactive.
A \cs \ cluster forms stable compounds when it is produced with
U, Ti, Zr, or Hf atoms trapped at its interior \cite{guo,dunlap,pedlao}.

There are three inequivalent sites on the \cs \ cluster:
we refer to atoms at the apex  sites as cage-A atoms, their immediate
neighbors as cage-B atoms, and the ones that are not connected by covalent
bonds to the cage-A atoms as cage-C atoms (see Figure~\ref{strucc28}).
It has been proposed that the \cs \ unit could be stabilized
by externally saturating the dangling bonds
of the four cage-A atoms
with hydrogen \cite{guo,haberlen,guo2,jkp}.
Saturation of the dangling bonds through intercluster covalent bonding has
also been considered:
The four cage-A atoms  of the \cs \ unit form the vertices
of a tetrahedron,
making the \cs \ unit analogous to a tetravalent atom.
A natural choice for a crystal composed of these units is the diamond
lattice \cite{guo,KLEIBY,p3}. This hypothetical solid,
called hyperdiamond
and symbolized by \cd,
is shown in Figure~\ref{struccd}(a) in a perspective view.
In Figure~\ref{struccd}(b), the same structure is
shown with all the cage-A, cage-B, and cage-C atoms on a single (110) plane
indicated.

A second candidate
which allows for
tetrahedral bonding of \cs \ units
is a compound of \cs \ clusters and tetravalent atoms in the
zincblende structure \cite{pedlao}.
Since the hyperdiamond lattice contains large voids in its structure,
one may expect that a solid composed of a more compact packing of
C$_{28}$ \ clusters alternating with individual
tetravalent atoms, would be energetically more favorable.
This kind of structure can be obtained by simply replacing the
central cluster in Figure~\ref{struccd} with a tetravalent atom and bringing
the neighboring clusters closer to the central atom to form covalent
bonds.
In this paper we consider a carbon atom as the second component of
the zincblende structure. We have performed extensive
first-principles calculations
to give a detailed comparison
between this zincblende structure and hyperdiamond. The
results are also compared to cubic diamond.

The rest of this paper is organized as follows:
Section~\ref{c28meth} describes the computational approach used in
our first principles calculations.
Section~\ref{energetics}  demonstrates the energetic
favorability of \csc \  over \cd \ through total energy
comparisons.  In
Section~\ref{elecstate} the electronic structure of
\csc \  is discussed in detail and compared
to that of diamond and of \cd.  Finally, in Section~\ref{summary}
we draw conclusions on the new \cs \ based solids and
comment on other possible solids based on similar cluster-atom combinations.

\section{Methods}
\label{c28meth}

We obtained total energies and
electronic densities, as well as single
particle electronic states and eigenvalues by
carrying out calculations
within the framework
of density functional theory and the local density approximation (DFT/LDA)
\cite{hohenberg,perdew}.
The optimal atomic coordinates and the equilibrium lattice constants are
obtained with a planewave basis, including plane waves with kinetic
energy up
to 36 Ry.
We also performed similar total energy
calculations for bulk diamond, in
order to make consistent total energy comparisons, and to obtain initial
estimates for equilibrium bond lengths in \csc.
For the reciprocal space integration, the
sampling  k-points have been chosen in a way such that
the density of k-points in the first Brillouin zone is kept approximately
constant for
all the structures we considered.
For bulk diamond, we used 125 k-points in the full Brillouin zone (which
corresponds to eleven special points in the irreducible Brillouin zone)
\cite{kpts},
with correspondingly smaller sets for the \csc \ and \cd \ structures.
Within the molecular dynamics
framework of Car and Parrinello \cite{carp}, we used the steepest
descent method for the initial relaxation of the electronic degrees of freedom,
followed by the more efficient conjugate gradient method close the
Born-Oppenheimer surface.

The ionic
potential, including the screening from core electrons,
was modeled by a nonlocal
norm-conserving pseudopotential \cite{bachelet},  and the
Kleinman-Bylander scheme \cite{kbl} was employed to make the
potential separable in Fourier space.
The $d$ angular momentum component
was treated as the local part of the
potential with the $s$ and $p$ components containing
the nonlocal contributions.

The choice of lattice constant for the \csc \ lattice
was guided by the
diamond calculations.  In \csc \ we first chose the length of
the bond between the single carbon atom and the C$_{28}$ \ cluster to
be the equilibrium bond length of diamond, as obtained by our
calculations.
Since the interaction between cage-B atoms of
neighboring C$_{28}$ \ clusters can affect the optimal length of this
C-C$_{28}$ bond, we performed the calculations at several different
lattice constants in the range near our initial choice,
corresponding to different values for the  C-C$_{28}$
bond length,
while the internal structure of the
\cs \ cage was held fixed at the one determined by
hydrogenation of the four cage-A atoms of the isolated cluster.  We used
this structure
because
the hydrogen atoms saturate the dangling bonds of \cs \
just as the single carbon atoms
do in \csc.
By fitting to a Birch-Murnaghan equation of state
\cite{BM}, we found that the lowest energy lattice constant corresponds to a
C-C$_{28}$ bond length of 1.53 \AA.  Fortuitously this was one of the actual
lattice constants for which the \csc \ calculations were performed.
For the optimal lattice so obtained, the positions
of the atoms were
then relaxed by minimizing the magnitude of forces obtained through the
Hellmann-Feynman theorem.
The cutoff below which the forces were considered to be negligible
is 0.01 Ry/a.u.
For the case of \csc , this relaxation not only
lowers the energy per atom but also
changes the electronic structure from a metal to a semiconductor, which
makes the crystal more stable. We discuss
this further in Section~\ref{elecstate}.
Similar calculation procedures were performed to obtain
the optimal structure in \cd.

\section{The Relative Total Energies}
\label{energetics}

The optimal total energy per atom
for \csc \ and
\cd \ after full relaxation are 0.453 eV and 0.744 eV respectively, where we
have used the energy per atom
of diamond as the reference.

The energy
of \csc \ is lower than that of hyperdiamond, which is due in part to the
presence of the
additional fourfold coordinated carbon atoms (the added single carbon atoms).
However, intercluster interactions also play an important role in stabilizing
\csc.
We can get an estimate of the  effects of intercluster interaction
by simply comparing the energy between
a unit cell
of \csc \ and the equivalent of twenty-eight atoms of \cd \ plus one atom of
diamond: the \csc \ lattice is lower in energy by 0.26 eV per atom.
This should be compared to the energy difference between \csc \ and \cd \
quoted above,
which is 0.291 eV.
Thus, intercluster interactions are more favorable in \csc,
rendering it  lower in energy than it
would be if the single added atoms were equivalent to diamond atoms
and the \cs \ clusters were equivalent to those of \cd.

We discuss next the effect of \cs \ cage relaxation on our results.
The relaxation has decreased the energy per atom for the case of \csc \
by as much as 0.24 eV/atom, while only approximately 0.01 eV/atom
is obtained through relaxation {\it from the same original choice of cage}
(that of the hydrogenated \cs \ cluster)
in the
case of \cd .
In order to understand this,
we first compare the distances between atoms in neighboring clusters between
different structures and how relaxation affects them.
In hyperdiamond, other than the covalent
bond between cage-A atoms on neighboring
clusters, the smallest distance between two atoms on neighboring
clusters is between a cage-A atom on one cluster and a cage-B atom on
the next cluster.
This distance is 2.57 \AA.  The next smallest distance between
atoms on neighboring clusters is between two cage-B
atoms and has a value of 3.15 \AA \ [see Figure 2(b)]. In this case, relaxation
does not
induce any significant change on the geometry of the cage structure
relative to the structure of the isolated, hydrogenated \cs \ cluster.

In contrast, in \csc \ the closest distance between atoms on
neighboring clusters before relaxation
is only 2.00 \AA, between two cage-B atoms of the original cages.
After the relaxation, this distance is reduced to 1.64 \AA, which
evidently introduces additional bonding between cage-B atoms as we will see
from the analysis of the electronic states in
Section~\ref{elecstate}.
Therefore, as far as the structural relaxation is concerned,
we have observed significantly different behavior
between \csc \ and \cd.
This was expected because the closest intercluster distance in the \csc \ solid
occurs
between two cage-B atoms, both of which are {\em threefold} coordinated.
An additional B-B intercluster bond will be energetically favorable
as carbon atoms prefer to be fourfold coordinated in this environment.
In the case of \cd \ the closest intercluster distance
is between a cage-B and a {\em fourfold} coordinated
cage-A atom, which makes the original cage structure in \cd \ electronically
more stable compared to \csc.

In Table~\ref{tabhd} we display the bond lengths of \csc \
and \cd \ before relaxation (the hydrogenated structure) and after relaxation.
The relaxation
in \cd \ produces
lengthening of the intercluster bonds
between cage-A atoms on neighboring \cs \ units and
shrinking of all other intracluster bonds.
The intracluster bonds which are shortened by the largest amount are those
between cage-C atoms,
which
are the farthest bonds
from the cage-A atoms.
In the case of \csc ,  a somewhat different relaxation pattern of the
cage geometry emerged: the C-C$_{28}$ bond and the C-C intracluster bonds
were shortened while the other two intracluster bonds were lengthened.
This relaxation can be attributed to the formation
of additional inter-cluster bonds between cage-B atoms.

\section {Electronic States}
\label{elecstate}
In Figure~\ref{tots} (a), (b), and (c),
we display the total valence electron densities
of the fully relaxed \csc \ and \cd \ solids and diamond
on the (110) plane. The  length scales in Figure~\ref{tots} (a)
and Figure~\ref{tots} (b) have been chosen so that these plots
cover approximately
equal areas.
It is evident from a comparison of these figures that \csc \ has a higher
atomic density than \cd.
Is is also apparent that
the basic \cs \ cluster geometry  remains
essentially unchanged in both the \csc \ and \cd \ structures.
This observation assures
us that the
\cs \ units are not altered significantly through the introduction of other
\cs \  clusters or additional carbon atoms in the solid forms.
The charge density around the additional carbon atom  outside the
cage (denoted by the symbol X in
Figure~\ref{tots}(a)) is
seen to be similar to the intercluster A-A bond in \cd.
These bonds are single covalent bonds
between two $fourfold$ coordinated carbon atoms:
comparison to
Figure ~\ref{tots}(c) shows that they have the same bonding
character as the bonds in bulk diamond.
The main difference between \cd \ and \csc \ is the presence of
additional bonds between  cage-B atoms of neighboring clusters
in the \csc \ crystal.
These bonds, seen clearly in Figure~\ref{tots}(a) between the two  cage-B atoms
of neighboring
clusters, are somewhat weaker than regular covalent bonds, as expected
from the fact that they are 1.64 \AA \ long, compared to the 1.54 \AA \ bond
distance in bulk diamond.

In Figure~\ref{dosc28} we display the density of states (DOS) of
\csc \ and \cd \ at the optimal lattice constants with full relaxation.
The DOS of diamond, as calculated in this work, is also shown in dashed lines
for comparison.
The DOS of the cluster-based solids exhibit many features that can be related
to features of the diamond DOS. For instance, the total
valence band width in all three cases is essentially the same, 21 eV.
Some specific features deserve closer attention:
In both cluster-based solids, the states corresponding to the
intercluster bonding between cage-A atoms, which is similar in nature to the
bonding in diamond, are far below the fermi level.
States immediately below and above the
fermi level derive from the
bonding properties  of
atoms within the clusters.
A careful examination
of the symmetry of the wave functions of the highest occupied and
lowest unoccupied states indicates that the states of \csc \
immediately below the fermi level are due to a combination of
intercluster bonding states
between cage-B atoms  and
$\pi$ bonding states between cage-C atoms within the cluster.
The same type of state involving intercluster bonding among
cage-B atoms is responsible for the highest occupied state in \cd,
even though the interaction is much weaker in that case.
On the other hand, the lowest unoccupied states in both solids
are primarily due to
a combination of intercluster bonding states between cage-B atoms and
antibonding
states between cage-C atoms within the cluster.

The DOS reveals that \cd \ is a
semiconductor with direct band gap,
which is consistent with results reported for \cd \ in
previous works \cite{guo,KLEIBY,p3}.
The fully relaxed \csc \ structure is also a semiconductor with
direct band gap equal to 1.16 eV,
approximately two thirds of the
band gap of \cd, and much smaller than the band gap of diamond.
As we have noted above, there is a much
stronger intercluster interaction in the case of \csc \ than in \cd;
it is therefore reasonable to expect that
the electrons should be more delocalized in \csc \ than in \cd,
which leads to the smaller band gap of \csc.
Here we wish to remind the reader of the well known inability
of DFT/LDA to reproduce the experimental band gaps in semiconductors
and insulators \cite{gaps}, so that all the numbers quoted above are
underestimates (by approximately a factor of 2) of the true band gaps,
if these solids would be realized.
For example, the band gap of crystalline cubic diamond
obtained by the present calculation
is 3.85 eV,
which is approximately 2/3 of the experimentally measured value 5.52 eV.

\section{Conclusions}
\label{summary}
In summary, we have studied two cluster-based solid structures,
the \csc \ and \cd \ lattices. Both structures satisfy the condition
that the dangling orbitals with $sp^{3}$ character on the cage-A
atoms are fully saturated.
Using first
principles calculations, we demonstrated that \csc \
is energetically preferred to \cd.
The crucial role of
intercluster interactions at the cage-B atoms  is
revealed by the structural relaxation in \csc, energetic comparisons, and the
valence electronic charge density.
The electronic structures of both \csc \ and \cd \
were analyzed. It was shown that for both solids
the intercluster bonding between cage-B atoms
and $\pi$ bonding between cage-C atoms within the cluster contribute to
the highest occupied states in \csc. The lowest unoccupied states, on
the other hand, come from a combination of the cage-B type intercluster bonding
states
mentioned above and anti-bonding states between cage-C atoms within the
cluster.

Just as intercluster interactions render \csc \ lower in energy
than \cd,
other solids based on a \cs \ unit with
saturated dangling bonds
might be candidates for stable structures as well.  For
example, we expect that solids analogous to \csc, but with the
single C atom that links the \cs \ units replaced by other
group-IV atoms (Si, Ge, Sn) would be equally stable, and with
similar electronic properties. Alternatively, solid forms composed of
\cs \ units connected by O atoms (chemical form C$_{28}$O$_{2}$) in a manner
analogous to silica or crystalline forms of SiO$_{2}$ could be rather
stable and somewhat easier to make, given the flexibility of packing of
tetrahedra bonded at their corners. The conditions under which such
solids may be experimentally realized remain to be investigated.

\section{Acknowledgements}
The first-principles calculations were carried out at the Pittsburgh
Supercomputer Center.  This work was supported in part by the Office of
Naval Research, contract N00014-92-J-1138.

\pagebreak
\vspace {9 mm}
\begin{table}[p]
\begin{center}
\begin{tabular}{||c|c|c|c|c|c||}  \hline
structure & lattice constant & bond A-B  &bond B-C&bond  C-C  &
C$_{28}$-C$_{28}$ \ bond \\
\cd \  & (\AA) & (\AA) & (\AA) & (\AA)& (\AA)   \\  \hline
unrelaxed & 15.777 & 1.525 & 1.412 & 1.507 & 1.504 \\
relaxed & 15.777  & 1.510 & 1.407 & 1.479 & 1.539 \\ \hline \hline
structure & lattice constant & bond A-B & bond B-C & bond C-C &
C-C$_{28}$ \ bond \\
\csc \  & (\AA) & (\AA) & (\AA) & (\AA)& (\AA)   \\  \hline
unrelaxed & 9.685 & 1.525 & 1.412 & 1.507 & 1.530 \\
relaxed & 9.685  & 1.574 & 1.469 & 1.435 & 1.501 \\ \hline
\end{tabular}
\end{center}
\caption{Bond lengths  before and after
relaxation of the cage structure for both the \cd \ and \csc \ structures.
The three types of intracluster bonds (bond A-B, B-C and C-C) refer to the
labels
of atoms shown in Figure 1 and Figure 2(b).
}\label{tabhd}
\end{table}
\dbs

\newpage
\newpage

\begin{figure}[p]
\caption[structet]{Structure of the \cs \ unit.  The dashed
lines indicate two of the four axes of $C_{3}$ symmetry;
the dashed-dotted line indicates one of the three axes of $C_{2}$ symmetry.
The three atom types are:
(i) an apex or cage-A atom , where
three 5-fold rings meet, (ii) a cage-B atom
which is part of
a 6-fold ring and is bonded to a cage-A atom, and (iii) a cage-C atom
which is also part of
a 6-fold ring, but is not bonded to a cage-A atom.
}\label{strucc28}
\end{figure}

\begin{figure}[p]
\caption[structet2]{
(a)  Perspective view of the structure of
the hyperdiamond lattice. The C$_{28}$\  clusters are bonded together
at the cage-A atoms shown here in black.
(b) The same structure as (a) shown along a (110) crystallographic
direction, where the positions of cage-A, cage-B, and cage-C atoms are
indicated
on a single plane.
The dashed line indicates the place where weak intercluster B-B
type bonding occurs in \cd.
The \csc \ structure is obtained by replacing the central cluster
by a single C atom and bringing the neighboring clusters closer to that site to
form covalent bonds.
In that case the interaction between neighboring cage-B atoms becomes much
stronger (see text).
}\label{struccd}
\end{figure}

\begin{figure}[p]
\caption[vale]{Total valence electron density
on the (110) plane of (a) \csc,
(b) \cd \ and (c) diamond.
The
white areas represent the highest electron density and the black areas
the lowest.
The positions of cage-A, cage-B and cage-C atoms are shown in (a) and (b).
The positions of the single extra carbon in \csc \ is indicated by the symbol
X;
the same symbol indicates the position of atoms in the diamond lattice (c).
}\label{tots}
\end{figure}

\begin{figure}[p]
\caption[dossol]{Density of states vs. energy for
the \csc \ (upper part) and \cd \ (lower part) structures.
For comparison the density of states of diamond (as calculated here)
is also shown in dashed
lines.
The fermi level (top of the valence
band) has been taken
to be the zero of the energy scale for each structure.
}\label{dosc28}
\end{figure}

\end{document}